\documentclass[showpacs,preprintnumbers,superscriptaddress,amsmath,amssymb,prl,twocolumn,floatfix]{revtex4}

\newcommand{\note}[1]{{\bf[#1]}}  
\renewcommand{\note}[1]{}  
\newcommand{\SKIP}[1]{}

\usepackage{graphicx}               
\usepackage{psfrag}
\usepackage{dcolumn}                
\usepackage{bm}                     
\usepackage{subfigure}

\newcommand{\ist}{\!=\!}

\renewcommand{\=}{\!=\!}
\newcommand{\gsim}{\raisebox{-3pt}{$\stackrel{>}{\sim}$}}

\newcommand{\bit}{\begin{itemize}}
\newcommand{\eit}{\end{itemize}}

\newcommand{\mean}[1]{{\big< {#1} \big> }}
\newcommand{\Eq}[1]{{Eq.~(\ref{#1})}}

\begin{document}

\title{Lattice Dynamics of the Heisenberg chain coupled to finite frequency bond phonons}

\author{Franz Michel}
\author{Hans Gerd Evertz}
\affiliation{Institute for Theoretical and Computational Physics,
   Graz University of Technology, Petersgasse 16, A-8010 Graz,
   Austria.}
\email{michel@itp.tugraz.at, evertz@tugraz.at}


\begin{abstract}
The phonon dynamics in a one dimensional Heisenberg spin chain coupled to 
finite-frequency bond phonons is studied. 
We present the first detailed phonon spectra for these systems using Quantum Monte Carlo.
The quantum phase transition is dominated by a central peak, 
yet the renormalisation of the main phonon branch depends strongly on the bare 
phonon frequency $\omega_0$.
The main branch remains largely unaffected at $\omega_0\gsim J$, but
it softens completely when $\omega_0$ is low enough. 
This is an unusual scenario for a structural phase transition
and was observable only on sufficiently large systems.
Approaching the dimerized phase from finite temperature, the lattice dynamics mirrors the 
behavior of a three dimensional system.
For the efficient measurement of Greens functions, 
we introduce a mapping from the stochastic series expansion to continuous imaginary time.
\end{abstract}


\maketitle
Due to the discovery of the inorganic spin-Peierls compound CuGeO$_3$, one dimensional spin systems 
coupled to phonons have been studied intensively with analytical and numerical methods like 
density-matrix-renormalisation \cite{BursillMH99}, exact diagonalisation \cite{WelleinFK98}, 
linked cluster expansion \cite{TrebstEM01}, continuous unitary transformation \cite{Uhrig98}, 
renormalisation group \cite{SunSB00}, and Quantum Monte Carlo (QMC) \cite{SandvikC99}.
Within a spin-Peierls transition, 
magnetic interactions cause a structural phase transition. It
can be classified into two categories: 
the displacive transition exhibits a soft phonon mode whereas the 
order-disorder transition is 
characterized by an additional central peak without any phonon softening \cite{BruceC81}.The 
Peierls-active phonon mode therefore possesses either one or two time scales. This, however, is a 
more phenomenological classification and even paradigm examples of displacive phase transition, 
like $SrTiO_3$ or $KMnF_3$, show an additional central peak above the critical temperature apart 
from the soft phonon mode \cite{RisteSOF71}. 
Within the spin-Peierls transition
there exist prominent candidates for either scenario: Copper bisdithiolene (TTF-CuBDT) 
has a precursive soft phonon mode at high temperature \cite{PRL39:507} 
and is therefore believed to belong to the soft phonon scenario. 
For CuGeO$_3$, however, although the central peak was not found yet directly, 
the Peierls active phonon mode hardens \cite{PRL80:3634} which is in agreement 
with the central peak scenario of the RPA 
\cite{CrossF79,PRB58:14677}.
Gros and Werner \cite{PRB58:14677} showed that RPA predicts softening only for bare phonon frequencies
smaller than some multiple of the transition temperature.

In this paper we focus on the phonon dynamics of the spin-1/2 anti-ferromagnetic Heisenberg chain 
interacting with finite frequency bond phonons.
This system  relates to  real materials as mentioned above, in which
the critical fluctuations are quasi one dimensional in a large temperature range above the critical 
temperature \cite{PougetRAH94}. 
Using sum rules, Sandvik and Campbell \cite{SandvikC99} have 
indirectly shown that the $q \= \pi$ phonon exhibits a central peak
even below the phase transition, for arbitrarily small spin-phonon 
coupling at $T\=0$. 
Therefore they suggested that the zero 
temperature transition is of the central peak type. 
We verify this conjecture by calculating the  whole phonon spectrum.
Contrary to the expectations of ref.\ \cite{SandvikC99}, we 
also find softening, and a 
qualitative dependence of the phonon spectra on the bare phonon frequency.
To our knowledge these are the first spectra for spin-Peierls transitions from unbiased QMC.
We also discuss the lattice dynamics at finite temperature.


The model under consideration is
\begin{equation}\label{eq:model}
    H \ist J\sum_{i=1}^N \left( 1 +g \ x_i\right) {\bf S}_i {\bf S}_{i+1} + \omega_0 \sum_{i=1}^N n_i
\end{equation}
where $
%
x_i = \frac{1}{\sqrt{2}} \left( a_i + a_i^\dagger \right)
%
$
is the spatial amplitude of the phonon which affects the lattice bond $(i,i+1)$, 
$g$ is the coupling between spins and phonons, $\omega_0$ is the bare 
frequency of the dispersion-less Einstein phonons (EP), 
and $n_i \= a_i^\dagger a_i$ is the phonon 
occupation number operator corresponding to $x_i$.
We use $J$ as our unit of energy.
The system is studied with a Quantum Monte Carlo (QMC) simulation based on the stochastic series 
expansion (SSE) \cite{SandvikSC97,SandvikK91}. The spins are updated globally using the loop algorithm 
\cite{EvertzLM93,Evertz03} whereas the phonons are updated locally.
%
To overcome the resulting high autocorrelations at large coupling $g$,
we have used simulated tempering \cite{MarinariP92} in 
$g$, with an automated choice of couplings and weights \cite{MichelE07}.

We study the phonon dynamics by analyzing the structure factor $S_x\left(q,\omega \right)$.
It is obtained via the inversion of 
\begin{equation} \label{eq:spectral-rep}
  \mean{x_{-q}\left( \tau \right)x_q\left(0\right)} 
  = \int\limits_{-\infty}^\infty d \omega ~S_x \left(q,\omega \right)~ e^{-\tau \omega},
\end{equation}
where $ x_q :\= \frac{1}{\sqrt{N}}  \sum_{j=1}^N  e^{- i q j} x_j$.

The measurement of such Greens functions 
in the SSE representation has been very cumbersome \cite{DorneichT01}.
Instead, we make use of a correspondence between SSE and continuous imaginary time 
\cite{SandvikSC97}. 
For a Hamiltonian $H\=H_0-V$, the interaction representation of the partition function is
\begin{equation}
 Z \!=\! \mbox{Tr} \sum_{n =0}^{\infty} 
        ~{\rm e}^{-\beta H_0} 
        \int_0^\beta \!\!\!d\tau_n \dots \int_0^{\tau_3} \!\!\!\!d\tau_2 \int_0^{\tau_2} \!\!\!\!d\tau_1 \,
          V(\tau_1 )\dots V(\tau_n) ,
\end{equation}
with  $V(\tau) \= {\rm e}^{ H_0\tau} V {\rm e}^{- H_0\tau}$.
Indeed, this is the representation of continuous time QMC 
\cite{ProkofevST96,Prokofev9698b,SandvikSC97,SandvikC99,SyljuasenS02,TroyerATW03}.
When $H_0\=0$, $V(\tau)$ does not depend on $\tau$, and the time integrals 
just provide a factor $\beta^n/n!$, which multiplies $V^n$.
When also $V\=H_1+H_2+\dots$, this product is a sum of ordered operator products,
which are the SSE operator strings.
Given an SSE configuration with $n$ operators, we can now stochastically map back to continuous time
by randomly choosing $n$ times between $0$ and $\beta$, ordering them,
and assigning them to the $n$ operators.
We obtain a configuration in space and continuous time. 
The Monte Carlo ensembles thus obtained in SSE representation and in imaginary time are equivalent.
Greens functions can then be measured efficiently 
by evaluating the measurements on a fine time-grid and using
the Fast Fourier Transform in space and time. 
Care should be taken with off-diagonal operators.
Details will be published elsewhere \cite{MichelE07}.
In our simulations, this procedure reduced the computational effort for measuring Greens functions
in all of momentum space by about a factor of $5 N$, where $N$ is the system size.

We improved the inversion of (\ref{eq:spectral-rep}) with the maximum entropy technique (MaxEnt) 
\cite{VonderLindenPH96} by using the exactly known moments
%
$ M_1 \= \frac {\omega_0} 2$ and $ M_3 \= \frac{{\omega_0}^3}{2}$, 
where $M_n \= \int \limits_{-\infty}^\infty d \omega  S\left(q,\omega \right) \omega^n$.
These moments 
roughly double the resolution of $S\left( q,\omega \right)$ near $\omega \sim \omega_0$.
The temperature in our simulations was chosen low enough to achieve the $T\=0$ limit where no changes in 
$S\left(q,\omega \right)$ by lowering the temperature are observed.
We used at least $\beta=2N$.



The spin-Peierls chain \Eq{eq:model} exhibits
a Kosterlitz-Thouless transition \cite{CitroOG05,SandvikC99}
at a finite coupling $g_c$ \cite{SandvikC99}
to a dimerized phase
with a finite spin gap $\Delta_S$, which begins, however, exponentially small in $g-g_c$.
Citro et al. \cite{CitroOG05} identified the relevance of the ratio of $\Delta_S$ to $\omega_0$:
at large $\Delta_S$ the system is adiabatic, 
while it is in an {\em anti}-adiabatic regime at small $\Delta_S/\omega_0$.
This is the case for finite frequency Einstein phonons until far beyond the phase transition.
Sandvik and Campbell \cite{SandvikC99} showed that for $0\<g\<g_c$ 
the phonon structure factor, 
which is $0.5$ at $g\=0$,
already diverges like $\log N$, 
reflecting a long range $1/r$ behavior in the 
phonon correlation function. 

We determined the critical coupling $g_c$ in the same way as ref.\ \cite{SandvikC99}
from the spin susceptibility. 
$\chi_s/N$ diverges for $g\<g_c$ 
(like ${\log}^{1/2} N$ for the pure spin model $g\=0$),
becomes constant at $g\=g_c$ where logarithmic corrections vanish,
and diminishes at larger $g$, reflecting the finite spin gap.
Our results for $g_c$ 
agree very well with recent DMRG calculations (Fig.\ 1 in ref.\ \onlinecite{WeisseHBF06}).

\begin{figure} [t]
  \centering
  \includegraphics[width=0.41\textwidth]{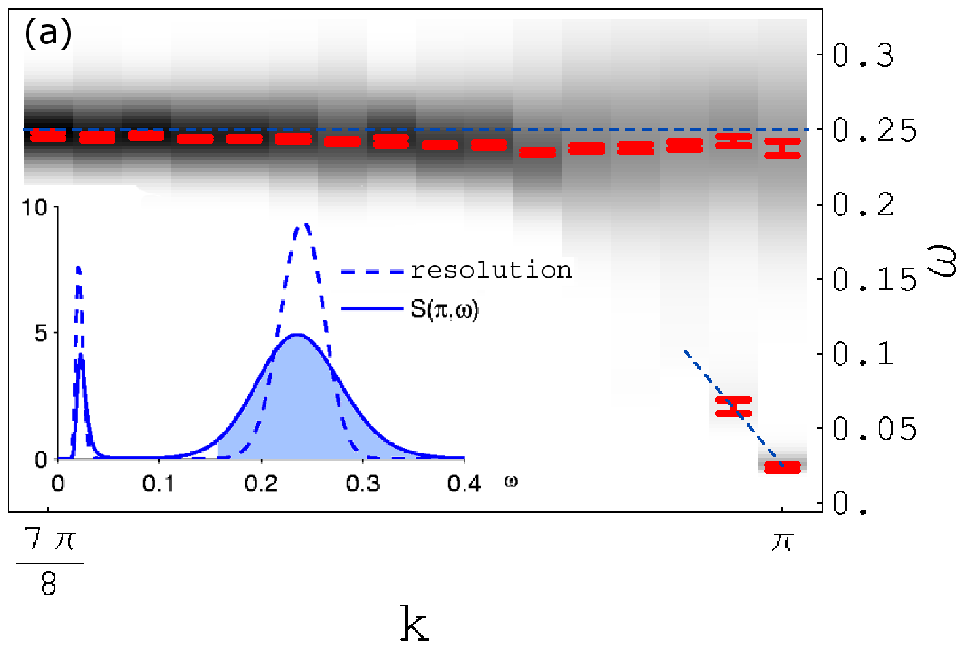}\\
  \includegraphics[width=0.41\textwidth]{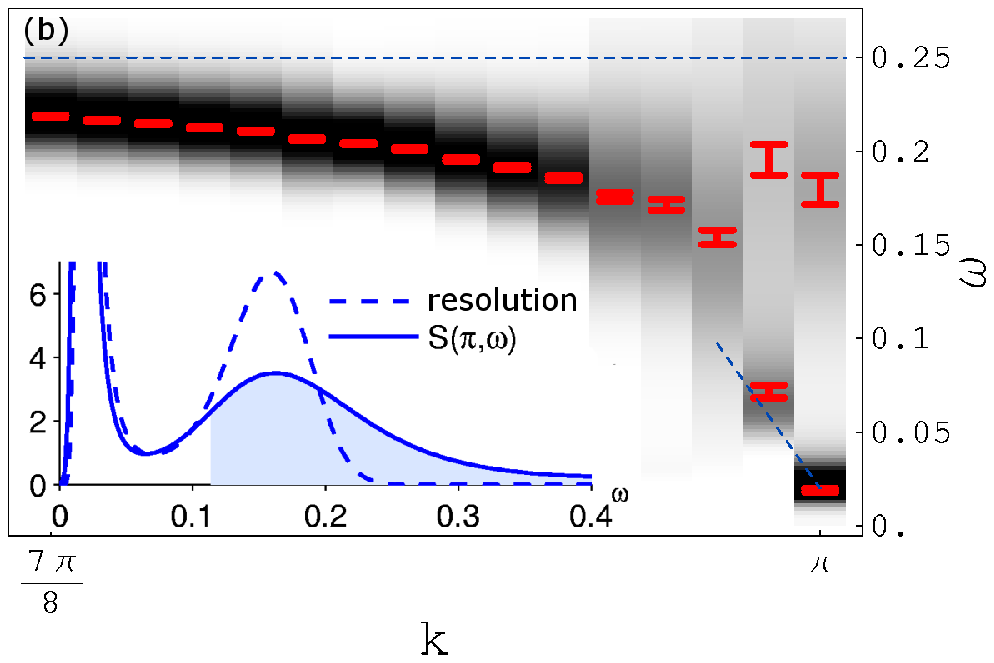}\\
   \includegraphics[width=0.43\textwidth]{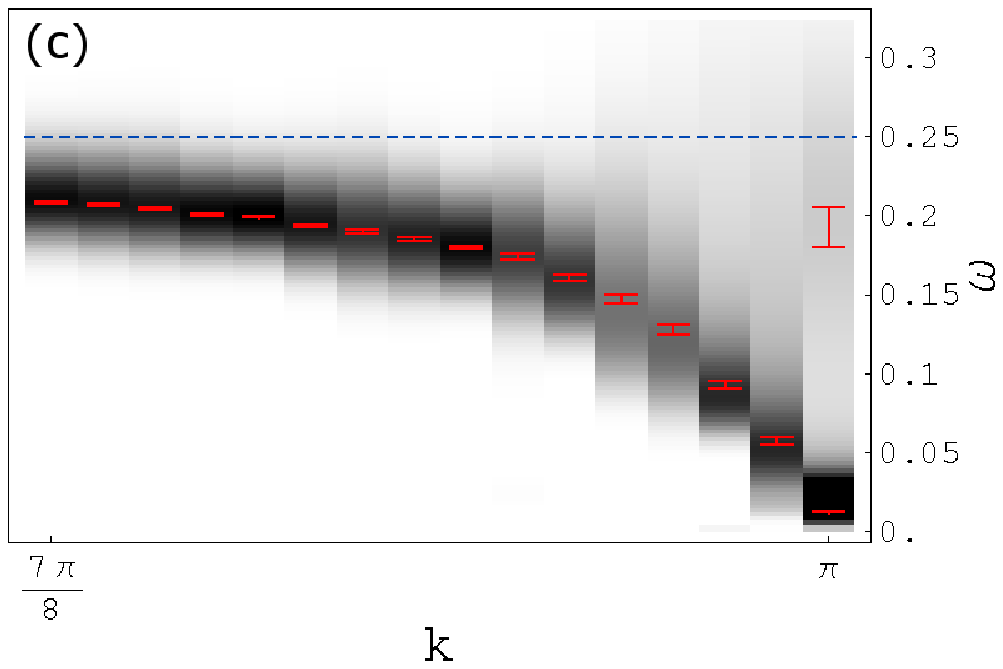}\\
    \caption{Dynamic structure factor $S_x(k,\omega)$ at small $\omega_0=0.25$ 
             in a density plot, 
             for spin-phonon couplings (a) $g=0.1$,  (b) $g=0.23$ close to the critical coupling, 
             and (c) $g=0.3$, at lattice size $N=256$ and $\beta=512$. 
             The error bars indicate the position of peaks and their uncertainties. 
             The insets show $S_x(\pi,\omega)$ and the MaxEnt resolution (see main text).
             The shaded blue area has a weight of 0.5, equal to the total spectral weight at $g=0$.
             For better visibility, the height of the low energy peak (central peak) at $k=\pi$ 
             has been truncated to the second highest peak in the spectra for $g=0.23$ and $0.3$.
             Depending on $g$, a central peak  as well as  softening occur.
}
    \label{fig:SpekOmega025}
\end{figure}

We will now discuss the phonon spectra.
They differ between small and large $\omega_0$.
We will first present results at small $\omega_0\=0.25$ (Fig.\ref{fig:SpekOmega025}).
Ref.\ \cite{SandvikC99} indirectly deduced the presence of a central peak
from the ratio of the phonon susceptibility and the static phonon structure factor.
The latter grows like $\log N$ up to the transition, and like $N$ beyond.

Indeed, at low coupling $g\=0.1$ (Fig.\ref{fig:SpekOmega025}a) we find a 
central peak and in addition a
new phonon branch starting close 
to $\omega\=0$ and $q\=\pi$ which is generated by the spin-phonon interaction. We will call 
it the central peak branch (CPB). The finite energy gap of the CPB at $q\=\pi$ scales to zero as 
$1/N$.
Away from $q\=\pi$ the spectral weight of the CPB decreases rapidly, leaving the EP well 
separated from the CPB. 
The dispersion of the CPB (dotted line) coincides with the spin velocity, as discussed later.

The insets of Fig.\ref{fig:SpekOmega025} provide some insight into the
distribution of spectral weights.
We use a test spectrum with two narrow peaks at the same position and with the same weight 
as $S_x(\pi,\omega)$. 
We add noise similar to that observed in $\mean{x_{-\pi}\left( \tau \right)x_\pi\left(0\right)}$
to the correlation function obtained from the test spectrum.
The resulting MaxEnt spectrum 
is shown as dashed lines. In the case of 
Fig.\ref{fig:SpekOmega025}a this shows that $S(\pi,\omega)$ is significantly wider than the MaxEnt 
resolution. The shaded area under $S(\pi,\omega)$ has a spectral weight of 0.5 which is the total 
spectral weight without spin-phonon interaction. The EP at $q\=\pi$ therefore widen but preserve 
their spectral weight.

\begin{figure} [t]
  \centering
\includegraphics[width=0.45\textwidth]{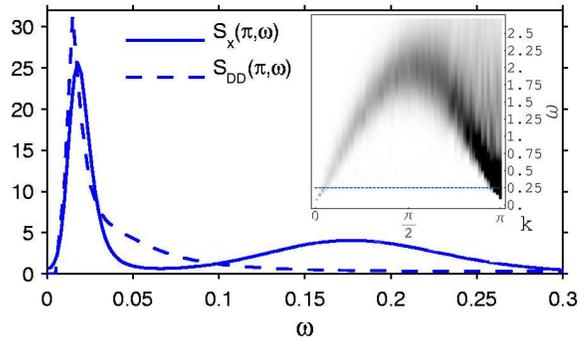}\\
    \caption{Comparison between the dynamical structure factor  
$S_{\bf{DD}}(\pi,\omega)$ of dimers  $\bf S_i \bf S_{i+1}$
and the phonon spectrum  $S_x (\pi,\omega)$ for $\omega_0 = 0.25$ and $g=0.23$,
at $N=256,\beta=512$.
Whereas the low energy peaks coincide, 
with a finite size energy gap which scales to zero like $1/N$,
$S_{\bf{DD}}(\pi,\omega)$ exhibits no structure at the soft phonon peak. 
The dimer spectrum here is almost identical to that of the pure spin model ($g=0$).
The inset shows the spin spectrum ($S_{S^zS^z}(\pi,\omega)$) as a function of $k$.
It reflects the effective increase of the spin coupling to about $1.1 J$ \cite{SandvikC99} 
by the phonon interaction, \Eq{eq:model}.
}
    \label{fig:DimerXXB512A023O025L256.eps}
\end{figure}

At  $g\=0.23 \simeq g_c$ (Fig.\ref{fig:SpekOmega025}b) 
the EP are strongly affected by the spin-phonon interaction. They soften partially.
The EP peak becomes broader and less well defined. 
Some spectral weight even extends to 
above $\omega \= 0.3$. 
The bare phonon frequency is not reflected in the spin spectra.
This can be seen in Fig.\ref{fig:DimerXXB512A023O025L256.eps} which shows the 
dimer-dimer structure factor in comparison to the phonon spectrum.
The spin, dimer, and phonon spectra exhibit a finite size gap.
It scales with increasing lattice size to a value undistinguishable from zero,
for all our calculations at $T=0$.

Beyond the phase transition, at  $g\=0.3$ (Fig.\ref{fig:SpekOmega025}c),
the spectrum is different.
The renormalized Einstein phonons have now joined the Central Peak Branch, 
forming a single new phonon branch,
which contains the complete diverging spectral weight.
The phonon spectrum is affected by the transition only for momenta fairly close to $\pi$.

\begin{figure} [t]
  \centering
  \includegraphics[width=0.45\textwidth]{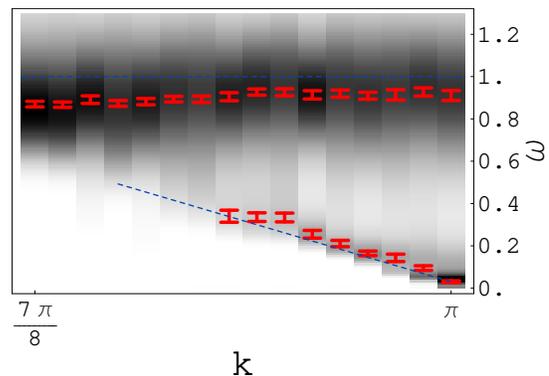}\\
    \caption{Dynamic structure factor $S_x(k,\omega)$  at large bare phonon frequency $\omega_0=1$
             for 
             $g=1$, close to but above 
             the critical coupling. ($N=256, \beta=512$).
             }
    \label{fig:omega1alpha05.eps}
\end{figure}

The situation is different for larger $\omega_0$. Fig. \ref{fig:omega1alpha05.eps} displays a 
density plot with $\omega_0\=1.0$ and $g\=1$ which is close to but above the critical coupling. 
There is a  central peak branch, which is now separated from the Einstein phonons, even above the transition.
No softening was observed at this larger $\omega_0$,
even at a large coupling of $g=1.6$.
Thus we find a qualitative difference between small $\omega_0$, where there is
a central peak, followed by softening beyond $g_c$,
and large $\omega_0$, which remains of a central peak nature.
%

%
The influence of spins on the phonons is directly visible in 
the dispersion relation of the central peak branch at $q \= \pi$, 
which is is the same as that of the dynamic dimer-dimer structure 
factor $S_{\bf{DD}}(q,\omega)$ and of the spinon branch in the spin spectrum within the accuracy of our 
simulations at all $\omega_0$.
The dotted lines in Fig.1 and Fig.2 depict a velocity of $v=\pi/2$,
which is the spin wave velocity of the pure Heisenberg chain.
This result is expected because the operator $x_i$ couples directly to the 
dimer-operator $\bf{S}_i\bf{S}_{i+1}$.

Since the system is critical for $g \le g_c$ we can also use the result of conformal field theory 
\cite{OkamotoN92} which predicts that the ground state energy $E_0(N)$ can be expressed as
\begin{equation} \label{eq:CFT}
E_0(N) =  
          E_0(\infty) 
          - \frac{\pi v}{6 N^2} \left[ 1+ \mbox{logarithmic corrections} \right].
\end{equation}
where $v$ is the velocity of the 
massless modes. Within our simulation we find 
the same velocity $v \= \pi/2 \pm 5 \%$ for $g  \le g_c$ and $\omega_0 \le 1$,
at large lattice sizes.
When extrapolating to the thermodynamic limit, 
we could not distinguish, within our resolution,
any logarithmic corrections nor a renormalization of the spin wave velocity 
occuring in the large $\omega_0$ limit \cite{FledderjohannG97}.

\begin{figure} [!!!!!tb]
  \centering
\includegraphics[width=0.45\textwidth]{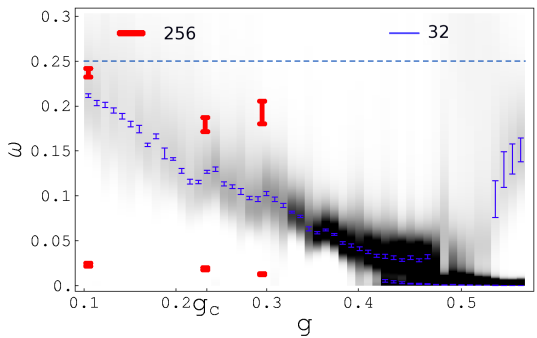}\\

 \caption{Comparison of the phonon spectra $S_x(\pi,\omega)$ at $q=\pi$ for different system sizes $N=32$ 
(blue) and $N=256$ (red) as a function of $g$ ($\omega_0=0.25$). While the bigger system exhibits a 
two peak structure with a central peak, the smaller system displays softening of the bare phonon 
mode.} 
    \label{fig:SoftHardfiniteSize.eps}
\end{figure}

\begin{figure} [!!!!!htb]
  \centering 
  \includegraphics[width=0.45\textwidth]{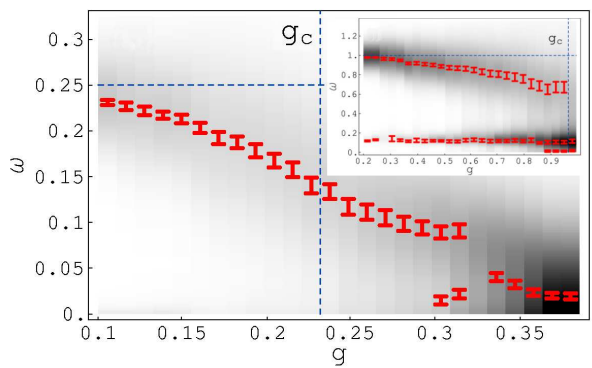}\\
    \caption{ $S_x\left(q=\pi ,\omega \right)$ as a function of spin-phonon coupling g for 
$\omega_0 = 0.25$  and $\omega_0 = 1$ (inset) at finite temperature $T=1/32$ (N = 128).}
    \label{fig:TemperatureA023O025L256.eps}
\end{figure}
Systems of small size behave in a qualitatively different way.
Instead of  two branches in $S_x(k\=\pi,\omega)$ 
they have only  a one peak structure.
The spectrum at $N\=32$ 
in Fig. \ref{fig:SoftHardfiniteSize.eps}
shows a softening of the $k\=\pi$ phonon mode
because of the finite size spin gap.
At $\omega_0=0.25$, large systems of $N\gsim 128$ 
are necessary to see the asymptotic two peak structure
and the new dispersing branch.

Similar behavior occurs at finite temperatures.
Fig.\ref{fig:TemperatureA023O025L256.eps} shows density plots of  $S_x\left(q\=\pi ,\omega \right)$ 
as a function of $g$ for $\omega_0 \= 0.25$ and $\omega_0 \= 1$ at $T\=1/32$. For  
$\omega_0 \= 1$ the situation is similar to the $T\=0$ limit. For  $\omega_0 \= 0.25$, however, there 
exits only one single phonon mode which softens with increasing coupling. At $g \sim 0.3$ we find a 
soft phonon mode accompanied by a CPB. 
Softening even without a CPB occurs for $\omega_0\=1$ at temperatures  $T\gsim 1/8$. 
Approaching the dimerized phase from finite temperature, 
the phonon dynamics 1D then resembles the 3D case,
where for a small ratio of $\omega_0/g$ a soft phonon mode is found, 
whereas for a bigger ratio 
the transition is governed by a central peak \cite{CrossF79}.

Thus we find the following lattice dynamics for the $T\=0$ transition in 1D,
with a qualitative dependence on $\omega_0$, which resembles the 3D behavior.
As soon as $g \neq 0$, 
a new narrow phonon branch emerges, centered at zero frequency and $q\=\pi$. 
It is clearly separated from the Einstein phonons at $\omega_0$. 
For big  $\omega_0$, the Einstein phonons are almost not affected by the spin-phonon interaction 
and stay separated from the central peak branch,
well beyond the phase transition.
For small $\omega_0$, the Einstein phonons soften significantly with increasing coupling 
before the transition takes place. 
For low enough  $\omega_0$ they soften so strongly that they eventually 
form a single branch with the central peak part.
Thus for low enough $\omega_0 $, the $T\=0$ structural phase transition presented here 
differs from its finite temperature counterpart in higher dimension 
in the sense that the phonons soften significantly even in the presence of a 
central peak.

To conclude, we have directly verified that the quantum phase transition to the dimerized state 
is of the central peak type. However, for low $\omega_0 \ll 1$ the Einstein phonons soften significantly 
and eventually join the central peak branch, forming a new phonon branch. 
%

{\it Acknowledgment}
This work has been supported by the Austrian Science Fund (FWF) projects P15520 and P15834. 
We would like to thank T.C. Lang and M. Daghofer for helpful discussions. 
The calculations made use of the ALPS libraries \cite{ALPS}.
\bibliography{spinpeierls_mod,qmc_mod}

\end{document}